\documentclass[prb,aps,twocolumn,showpacs,nofootinbib]{revtex4}

\usepackage{graphicx}
\usepackage{rotating}
\usepackage{amsmath}
\usepackage{amsfonts}
\usepackage{amssymb}
\usepackage[countmax]{subfloat}
\usepackage{dcolumn} 
\usepackage{bm} 
\usepackage{color}

\usepackage{float}
\usepackage{latexsym}

\begin{document}

\title{Flux Quantization Due to Monopole and Dipole Currents}

\author{Wei Chen$^{1}$, Peter Horsch$^{1}$, and Dirk Manske$^{1,2}$}

\affiliation{$^{1}$Max-Planck-Institut f$\ddot{u}$r Festk$\ddot{o}$rperforschung, Heisenbergstrasse 1, D-70569 Stuttgart, Germany
\\
$^{2}$Yukawa Institute for Theoretical Physics, Kyoto University, Kyoto 606-8502, Japan}

\date{\rm\today}

\begin{abstract}

{By discussing field-induced quantum interference effects due to monopole moments and those due to dipole moments on equal footing, their similarities and differences are clarified. First, we demonstrate the general principle for flux quantization. For particles carrying a monopole moment, the interference causes monopole current to oscillate periodically with flux defined as inner product of field and area, whereas for particles carrying a fixed dipole moment, the dipole current oscillates periodically with flux vector defined as cross product of field and trajectory. Our analysis unifies the oscillation of monopole or dipole currents in various devices, such as SQUID and spin-FET, into the same physical picture.  Second, we show that interference effects can also happen in open trajectory devices that transport dipole currents, such as spin Josephson effect, based on the non-gauge field nature of the interference effects of dipole moments. In addition, we propose that the interference effect of electric dipoles, known as He-McKellar-Wilkens effect, can be realized by the bilayer exciton condensates observed in semiconductor heterostructure and bilayer graphene.  }

\end{abstract}







\maketitle

\section{Introduction}

Quantum interference effects due to external electric or magnetic fields have long been of great interest, since they demonstrate the feasibility of controlling the quantum state of particles by external fields and preserving quantum coherence. Such interference effects stem from the coupling of monopole or dipole moment of the particles to the vector field or the electric and magnetic field via Aharonov-Bohm (AB)\cite{Aharonov59}, dual Aharonov-Bohm (DAB)\cite{Dowling99}, Aharonov-Casher (AC)\cite{Aharonov84}, and He-McKellar-Wilkens (HMW)\cite{He93,Wilkens94} effects. Due to these effects, the wave function of the particle acquires a Berry phase\cite{Berry84} after traveling along a certain trajectory, and allows to utilize the Berry phase to design various interferometers.

Since the Berry phase is a periodic argument, a natural consequence of field-induced interference effects is the flux quantization\cite{Deaver61,Doll61}. Application of magnetic flux quantization in solid state devices includes superconducting quantum interference devices (SQUIDs), in which the Berry phase due to AB effect combined with single-valuedness of the wave function yields a Josephson current oscillating periodically with magnetic flux, from which the flux quantization is interpreted. The magnetic flux is defined as the inner product of magnetic field and the area enclosed by the trajectory; and the smallness of the flux quantum $h/2e$ is the reason behind the high precision magnetometer made by SQUID.

In this article, we show that flux quantization can also be introduced in DAB, AC, and HMW effects in the same sense as flux quantization in a SQUID, i.e., from the oscillation of monopole or dipole currents with flux in solid state devices. We follow the principle that the flux should be defined as the quantity that controls the interference effect but only depends on the field and the trajectory, whereas the monopole or dipole moments of the particles determine the flux quantum. This principle unambiguously yields a scalar for fluxes associated with monopole moments, but a vector for fluxes associated with fixed 
magnetic or electric
dipole moments. We show that this general picture applies to a great number of devices in many different fields, including spintronic\cite{Wolf01,Zutic04,Fabian07} and excitonic\cite{Rashba82,Scholes06} systems, and provides a unified picture for all known field-induced interference effects for nonrelativistic particles with fixed monopole or dipole moments. This unified picture also motivates us to seek analog of devices that use AB effect in other interference effects. In particular, we show that the long sought HMW effect can be realized by bilayer exciton condensates proposed\cite{Lozovik75,Kuramoto78,Fertig89,Zhang08,Min08,Kharitonov10} and observed recently in semiconductor heterostructures\cite{Spielman00,Kellogg04,Tutuc04,Eisenstein04,Tiemann08,Huang12,Sivan92,Seamons07,DasGupta08} and bilayer graphene\cite{Kim11,Kim12,Gorbachev12,Berman12}. A dc SQUID-like device is suggested to observe the interference of the exciton condensate. This device brings quantum Hall systems and graphene into applications in quantum interference, and may in turn be used to determine the experimental value of the electric dipole moment of bilayer excitons.

In addition, by comparing the interference effects due to monopole moments and that due to dipole moments, we recognize the importance of the non-gauge field nature of the interference effects involving point dipole moments, hence the possibility to observe them in open trajectory devices. As an example, we propose a single trajectory interferometer based on spin Josephson effect\cite{Lee03,Nogueira04,Asano06,Linder07,Brydon09,Chasse10,Moor12}, which acts as a $\varphi$-junction where the phase of the Josephson spin current can be arbitrarily controlled by a gate voltage. This example demonstrates that interference in open trajectories is a new concept that allows for design of new interferometers.

The structure of the paper is the following. In Sec. II, we discuss a generic setup that demonstrates the flux quantization for all four effects considered. Sec. III addresses the observability of quantization of electric flux vector. An open trajectory spin interferometer is proposed in Sec. IV, and a dc SQUID-like device is proposed in Sec. V to observe the interference of bilayer exciton condensates. Sec. VII gives a summary of the results.

\section{General principle of flux quantization}

\begin{figure}[ht]
\begin{center}
\includegraphics[clip=true,width=0.9\columnwidth]{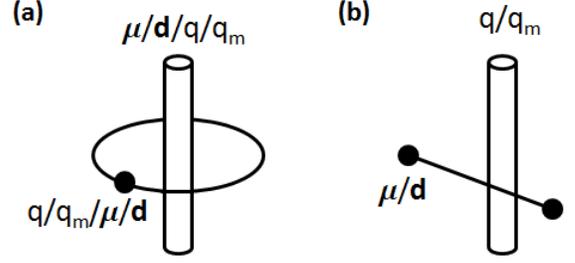}
\caption{ (color online) (a) Generic setup that manifests oscillation of monopole or dipole currents due to AB, DAB, AC, and HMW effects, which consists of $q/q_{m}/{\boldsymbol\mu}/{\bf d}$ in a mesoscopic ring that experiences the field from uniformly distributed ${\boldsymbol\mu}/{\bf d}/q/q_{m}$ on a line pierced through the ring, respectively. (b) Schematics of the open trajectory devices in which AC and HMW effects can take place. } 
\label{fig:demonstrations}
\end{center}
\end{figure}

We first demonstrate flux quantization that stems from the oscillation of monopole or dipole currents as a result of quantum interference. This is best demonstrated within the framework of persistent current in a mesoscopic ring\cite{Byers61,Buttiker83}, in combination with the generic setup proposed in Ref.~\onlinecite{Dowling99} where monopole moments and dipole moments can be discussed on equal footing, as shown in Fig. \ref{fig:demonstrations}(a). Quantum particles carrying electric charge ($q$)/magnetic monopole ($q_{m}$)/ magnetic dipole (${\boldsymbol\mu}$)/electric dipole (${\bf d}$) are confined in a 1D ring of length $L$, with an infinitely long wire carrying uniformly distributed ${\boldsymbol\mu}$/${\bf d}$/$q$/$q_{m}$ pierced through the center, which corresponds to the setup for AB/DAB/AC/HMW effects, respectively. First we review the mechanism of persistent charge current due to AB effect in this setup, then we make analogy to the other three effects.

In the setup for AB effect, the momentum ${\bf p}$ of a particle with electric charge $q$ is replaced by (SI units are adopted throughout the article, and boldface symbols denote vectors)
\begin{eqnarray}
AB:\;\;m{\bf v}={\bf p}-q{\bf A}\;\rightarrow\;\varphi_{AB}=\frac{q}{\hbar}\oint{\bf A}\cdot d{\bf l}=2\pi\frac{\Phi_{B}}{\Phi_{B}^{0}} ,
\label{momentum_AB}
\end{eqnarray}
that is, from the form of momentum, the Berry phase $\varphi_{AB}$ is already determined. The discrete eigenenergies for the particles with mass $m$ are 
\begin{eqnarray}
E_{n}=\frac{2\pi^{2}\hbar^{2}}{mL^{2}}\left(n-\Phi_{B}/\Phi_{B}^{0}\right)^{2}.
\label{eg_energies}
\end{eqnarray}
The many particle energy spectrum is periodic in $\Phi_{B}=\oint{\bf A}\cdot d{\bf l}=\int{\bf B}\cdot d{\bf a}$ with periodicity $\Phi_{B}^{0}=h/q$, and so is the charge current calculated by $I=\sum_{n}\left(q/\hbar L\right)f(E_{n})\partial E_{n}/\partial k$, where $f(E_{n})$ is the Fermi or Bose distribution function depending on the statistics of the particles\cite{Cheung88}. We emphasize that although the  AB effect is due to coupling of $q$ to the vector field ${\bf A}$, the magnetic flux is defined only through the identity $\oint{\bf A}\cdot d{\bf l}=\int{\bf B}\cdot d{\bf a}$.

Now consider the setup for the DAB effect. The electric field due to ${\bf d}$ distributed on the wire is written in terms of a vector field\cite{Dowling99} ${\bf E}={\boldsymbol\nabla}\times{\bf A}_{E}$, and the momentum of the particle that carries $q_{m}$ is replaced by
\begin{eqnarray}
DAB&:&\;\;\;m{\bf v}={\bf p}+\frac{q_{m}}{c^{2}}{\bf A}_{E}\;
\nonumber \\
&\rightarrow&\;\varphi_{DAB}=-\frac{q_{m}}{\hbar c^{2}}\oint{\bf A}_{E}\cdot d{\bf l}=2\pi\frac{\Phi_{E}}{\Phi_{E}^{0}}\;,
\label{momentum_DAB}
\end{eqnarray}
where $\Phi_{E}=-\oint{\bf A}_{E}\cdot d{\bf l}=-\int {\bf E}\cdot d{\bf a}$, and $\Phi_{E}^{0}=hc^{2}/q_{m}$. The eigenenergies have the same form as Eq.~(\ref{eg_energies}), with the replacement $\Phi_{B}/\Phi_{B}^{0}\rightarrow\Phi_{E}/\Phi_{E}^{0}$. Following the same argument as in the AB effect, both the eigenenergies and monopole current are periodic in $\Phi_{E}$ with periodicity $\Phi_{E}^{0}$, hence the quantization of $\Phi_{E}$ is interpreted.

It is intriguing to ask if our analysis can be applied to the recently discovered magnetic monopoles in spin ice \cite{Castelnovo08,Morris09,Bramwell09,Ryzhkin05} or quantum spin ice \cite{Hermele04,Savary12}. For instance, if piercing an ${\bf E}$ field through a ring made of quantum spin ice can generate a monopole current that oscillates with $\Phi_{E}$. However, it is clear that in our analysis, we assume an ideal monopole whose presence modifies Maxwell's equation, and it experiences a Lorentz force in the presence of an ${\bf E}$ field. Since the monopoles in quantum spin ice do not satisfy such criteria, the DAB effect does not occur in these materials. Therefore our analysis for such ideal monopoles is rather for the sake of completing the duality between magnetism and electricity, whereas its realization depends on the existence of ideal monopoles.

 For AC and HMW effect of charge neutral dipoles, the Lagrangians are\cite{Aharonov84,Wilkens94} $L_{AC}=m{\bf v}^{2}/2+{\bf v}\cdot\left({\boldsymbol\mu}\times{\bf E}\right)/c^{2}$ and $L_{HMW}=m{\bf v}^{2}/2+{\bf v}\cdot\left({\bf B}\times{\bf d}\right)$. From ${\bf p}=\partial L/\partial{\bf v}$, one deduces the relation between velocity and canonical momentum 
\begin{subequations}
\begin{eqnarray}
&&AC:\;m{\bf v}={\bf p}-\frac{1}{c^{2}}{\boldsymbol \mu}\times{\bf E}\;,
\\
&&HMW:\;\;m{\bf v}={\bf p}-{\bf B}\times{\bf d}\;.
\label{momentum_AC_HMW}
\end{eqnarray}
\end{subequations}
By comparing with Eq. (\ref{momentum_AB}), the Berry phase acquired by the dipole is
\begin{subequations}
\begin{eqnarray}
&&\varphi_{AC}=\frac{1}{\hbar c^{2}}\int\left({\boldsymbol \mu}\times{\bf E}\right)\cdot d{\bf l}
=\frac{|{\boldsymbol \mu}|}{\hbar c^{2}}{\boldsymbol{\hat\mu}}\cdot\left(\int {\bf E}\times d{\bf l}\right)
\nonumber \\
&&=2\pi\frac{{\boldsymbol{\hat\mu}}\cdot{\boldsymbol \Phi}_{E}}{\tilde{\Phi}_{E}^{0}} ,
\\
&&\varphi_{HMW}=\frac{1}{\hbar}\int\left({\bf B}\times{\bf d}\right)\cdot d{\bf l}=\frac{|{\bf d}|}{\hbar}{\bf{\hat d}}\cdot\left(\int d{\bf l}\times{\bf B}\right)
\nonumber \\
&&=2\pi\frac{{\bf{\hat d}}\cdot{\boldsymbol \Phi}_{B}}{\tilde{\Phi}_{B}^{0}} .
\label{phase_AC_HMW}
\end{eqnarray}
\end{subequations}
Here we confine our discussion to cases where the dipole moments 
do not vary
along their trajectory. By factoring the Berry phase into the part that depends only on the field and trajectory and the part that depends on the fixed dipole moment, the flux is unavoidably a vector defined by the cross product of field and  trajectory. Below we call ${\boldsymbol\Phi}_{E}$ the electric flux vector, and ${\boldsymbol\Phi}_{B}$ the magnetic flux vector. The eigenenergies follow Eq.~(\ref{eg_energies}), with the replacement $\Phi_{B}/\Phi_{B}^{0}\rightarrow{\boldsymbol{\hat\mu}}\cdot{\boldsymbol\Phi_{E}}/\tilde{\Phi}_{E}^{0}$ and $\Phi_{B}/\Phi_{B}^{0}\rightarrow{\bf{\hat d}}\cdot{\boldsymbol\Phi_{B}}/\tilde{\Phi}_{B}^{0}$ for AC and HMW effect, respectively. Both the eigenenergies and dipole currents are periodic in ${\boldsymbol\Phi}_{E}$ and ${\boldsymbol\Phi}_{B}$, from which their quantization follows. The actual quantized values of ${\boldsymbol\Phi}_{E}$ and ${\boldsymbol
 \Phi}_{B}$ depend not only on their flux quanta $\tilde{\Phi}_{E}^{0}=hc^{2}/|{\boldsymbol\mu}|$ and $\tilde{\Phi}_{B}^{0}=h/|{\bf d}|$, but also on the direction of the fixed dipole moments.

The oscillation of dipole currents with field has been discussed by Balatsky and Altshuler, who considered persistent spin currents in a $^{3}$He ring\cite{Balatsky93}, and the electric flux quantum $\tilde{\Phi}_{E}^{0}$ has been introduced by Bogachek and Landman\cite{Bogachek94}. While these considerations refer to flux quantization in closed circuits, our point is that this oscillation obeys a general principle of flux quantization where the proper definition of flux is a vector for fixed dipole moments. Some devices may have more than one of the above effects, for instance the $^{3}$He ring\cite{Balatsky93} and a spin filter/reader proposed recently\cite{Aharony11}, in which both AB and AC effect contribute to interference. Flux quantization can still be interpreted in these devices as long as one external field controls only one interference effect, for instance ${\bf E}$ controls $\varphi_{AC}$ linearly, and causes the current to oscillate periodically.
  In the following sections, we address how quantization of flux vectors can be realized in concrete devices.


\begin{table}
\begin{tabular}{ p{2.5cm}<{\centering}  p{2.5cm}<{\centering}  p{2.5cm}<{\centering}  }
\hline \hline

     & electric & magnetic \\ [0.5ex] \hline 

& AB & DAB \\
monopole & $\Phi_{B}=\int  d{\bf a}\cdot{\bf B}$, $\Phi_{B}^{0}=h/q$ & $\Phi_{E}=-\int d{\bf a}\cdot{\bf E}$, $\Phi_{E}^{0}=hc^{2}/q_{m}$  \\ [0.5ex] \hline 
& HMW & AC \\
dipole & ${\boldsymbol\Phi}_{B}=\int d{\bf l}\times{\bf B}$, ${\tilde\Phi}_{B}^{0}=h/|{\bf d}|$ & ${\boldsymbol\Phi}_{E}=-\int d{\bf l}\times{\bf E}$, ${\tilde\Phi}_{E}^{0}=hc^{2}/|{\boldsymbol\mu}|$ \\ [0.5ex] \hline 
\hline
\end{tabular}	
\caption{ 
Definition of flux and flux quantum in the four field-induced interference effects under discussion, classified according to the fixed electric or magnetic monopole or dipole moment of the particles.}
\label{tbl:summary_flux_quantization}
\end{table}

\section{On the observation of quantization of electric flux vector}

For particles that carry electron magnetic moment $\mu_{B}$ and experience AC effect, the quantum for electric flux vector ${\boldsymbol\Phi}_{E}$ is huge ${\tilde\Phi}_{E}^{0}=hc^{2}/\mu_{B}=6.43\times 10^{6}$V. This means for a typical ${\bf E}$ field in the laboratory, $\varphi_{AC}$ is only of the order of mrad\cite{Cimmino89}. In comparison with the smallness of magnetic flux quantum $\Phi_{B}^{0}=h/2e=2.07\times 10^{-15}$Wb, one may wonder if such a huge flux quantum and the resulting small phase shift can be of any use. Below we demonstrate that in devices that contain Rashba spin-orbit coupling (SOC), which yields a phase shift also proportional to ${\boldsymbol\mu}\times{\bf E}$ but with a much larger prefactor, the flux quantum may be experimentally observable.

A prototype interference device that utilizes Rashba SOC is the spin field effect transistor (spin-FET)\cite{Datta90}. In an ideal 1D spin-FET with length $L$ along $x$-direction, the spin degeneracy ($\sigma=\pm$) is lifted in the 2DEG region where an ${\bf E}$ field is applied along ${\bf \hat{y}}$ direction, described by $E_{\sigma}=\hbar^{2}k_{x\sigma}^{2}/2m^{\ast}-\sigma\alpha k_{x\sigma}$, which corresponds to the Hamiltonian 
\begin{eqnarray}
H_{\sigma}&=&\frac{1}{2m}|{\bf p}_{\sigma}-\sigma{\boldsymbol\mu}\times\left({\boldsymbol g}+\lambda{\bf E}\right)|^{2}
-\frac{1}{2m}|{\boldsymbol\mu}\times\left({\boldsymbol g}+\lambda{\bf E}\right)|^{2}\;.
\nonumber \\
\end{eqnarray} 
The Rashba coupling corresponds to $\alpha=\hbar|{\boldsymbol\mu}\times\left({\boldsymbol g}+\lambda{\bf E}\right)|/m$, where ${\boldsymbol g}$ represents the intrinsic SOC of the 2DEG, and $\lambda$ characterizes the field induced SOC. It is still debated whether the field dependence of $\alpha$ comes from the expectation value of the electric field at the interface\cite{Nitta97}, or the asymmetry of the wave function in the quantum well\cite{Engels97,Zawadzki04}, or other aspects of the wave function\cite{Grundler00}. For either mechanism to be true, our point is that the Rashba parameter can be {\it empirically} written in the form $\alpha=\hbar|{\boldsymbol\mu}\times\left({\boldsymbol g}+\lambda{\bf E}\right)|/m$ where ${\bf E}$ is the external field depending only on the gate voltage and sample thickness. By preparing the spins in the source and drain in an eigenstate of $\sigma_{x}$, the tunneling probability, 
\begin{eqnarray}
P_{0}\propto 1+\cos\Delta\theta\;,
\end{eqnarray}
oscillates with 
\begin{eqnarray}
\Delta\theta=\int\left({\bf k}_{+}-{\bf k}_{-}\right)\cdot d{\bf l}=\left(k_{x+}-k_{x-}\right)L=\frac{2m\eta L}{\hbar^{2}}\;,
\end{eqnarray}
which can be rewritten as
\begin{eqnarray}
\Delta\theta&=&\frac{2|{\boldsymbol\mu}|}{\hbar}{\boldsymbol{\hat\mu}}\cdot\left(\int {\boldsymbol g}\times d{\bf l}\right)+\frac{2\lambda|{\boldsymbol\mu}|}{\hbar}{\boldsymbol{\hat\mu}}\cdot\left(\int {\bf E}\times d{\bf l}\right)
\nonumber \\
&=&\varphi_{0}+2\pi\frac{{\boldsymbol{\hat\mu}}\cdot{\boldsymbol\Phi}_{E}}{{\tilde\Phi}_{E}^{0}}\;.
\end{eqnarray}
Therefore from the periodicity of tunneling probability, or equivalently the current-voltage characteristics, the quantization of ${\boldsymbol\Phi}_{E}$ is realized, with flux quantum ${\tilde\Phi}_{E}^{0}=h/2\lambda|{\boldsymbol\mu}|$. Physically, the quantization of ${\boldsymbol\Phi}_{E}$ shows up because spin-FET utilizes the Berry phase of spin up and down (which are fixed magnetic moments) to control the tunneling probability, so it falls into the category of field-induced quantum interference effect. Applying a gate voltage $\sim 1$V on an inverted In$_{0.53}$Ga$_{0.47}$As/In$_{0.52}$Al$_{0.48}$As heterostructure of thickness $\sim 100$nm and channel length $\sim\mu$m can obtain $\Delta\theta=2\pi$\cite{Nitta97}, which corresponds to a flux quantum $\tilde{\Phi}_{E}^{0}\sim 10$V. Thus the flux quantum due to Rashba SOC is generally much smaller and more accessible compared to that in AC effect. One also sees that the flux quantum may be used to determine the parameter $\lambda$.

\section{Flux quantization in open trajectory devices}

Although we discuss the flux quantization due to monopole currents and due to dipole currents on equal footing, a crucial difference between them should be emphasized. The quantization due to monopole current originates from the coupling of monopole moment and the vector field ${\bf A}$ and ${\bf A}_{E}$ described in Eqs. (\ref{momentum_AB}) and (\ref{momentum_DAB}), which can always be gauged away unless the particle is moved in a closed trajectory, such that the gauge invariant fluxes $\oint{\bf A}\cdot d{\bf l}=\int{\bf B}\cdot d{\bf a}$ and $-\oint{\bf A}_{E}\cdot d{\bf l}=-\int{\bf E}\cdot d{\bf a}$ can be measured. On the other hand, the quantization due to dipole current stems from the coupling of dipole moment and the external field directly, which cannot be gauged away even if the particle moves in an open trajectory, as demonstrated in Fig.~\ref{fig:demonstrations}(b). In other words, AC and HMW effects can cause oscillation of dipole current in open trajectory devices.

\begin{figure}[ht]
\begin{center}
\includegraphics[clip=true,width=0.9\columnwidth]{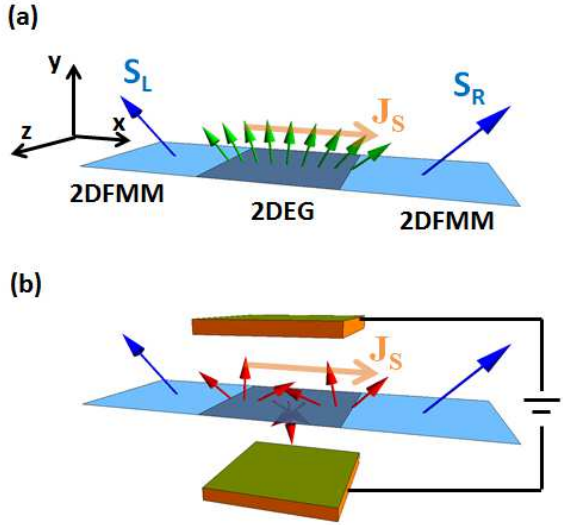}
\caption{ (color online) (a)Proposed 2DFMM/2DEG/2DFMM junction as a realization of the open trajectory device sketched in Fig. \ref{fig:demonstrations}(b). Magnetization $\langle{\bf S}_{R}\rangle$ and $\langle{\bf S}_{L}\rangle$(blue arrows) yield the Josephson spin current $J_{s}$(polarization ${\boldsymbol\mu}\parallel\langle{\bf S}_{R}\times{\bf S}_{L}\rangle$) described by Eq. (\ref{Js_FET}). The magnetization profile induced at the 2DEG interface(green arrows) gradually rotates from $\langle{\bf S}_{R}\rangle$ to $\langle{\bf S}_{L}\rangle$. (b)Applying a gate voltage yields a phase shift in the Josephson spin current. The induced magnetization profile(red arrows) changes accordingly, which should be visible as certain fringe pattern by polarization-sensitive probes. } 
\label{fig:2DFMM2DEG2DFMM}
\end{center}
\end{figure}

An example of such open trajectory devices is the Josephson junction. This motivates us to study spin Josephson effect in the presence of an electric field. In particular, we revisit the spin Josephson effect due to coherent tunneling of spinful particle-hole pair $\langle c_{\uparrow}^{\dag}c_{\downarrow}\rangle$ in a ferromagnetic metal/insulator/ferromagnetic metal(FMM/I/FMM) junction\cite{Nogueira04}. Comparing to other types of Josephson junctions that manifest tunneling of Cooper pairs\cite{Asano06}, the charge neutral $\langle c_{\uparrow}^{\dag}c_{\downarrow}\rangle$ couples to external electric field via SOC or AC effect, and there is no Josephson charge current in this problem. We follow the analysis in Ref.~\onlinecite{Nogueira04} for FMM/I/FMM junction but with a different mean field treatment of Hubbard interaction in the bulk FMM
\begin{eqnarray}
Un_{\uparrow}({\bf r})n_{\downarrow}({\bf r})
\rightarrow-c_{\uparrow}^{\dag}({\bf r})c_{\downarrow}({\bf r})\Delta({\bf r})-c_{\downarrow}^{\dag}({\bf r})c_{\uparrow}({\bf r})\Delta({\bf r})^{\dag},
\label{mean_field}
\end{eqnarray}
where $\Delta^{\dag}=U\langle S^{+}\rangle=U\langle S^{x}+iS^{y}\rangle=U\langle c_{\uparrow}^{\dag}c_{\downarrow}\rangle$. In this treatment, the magnetization lies in the $S^{x}S^{y}$-plane with an angle $\theta$ that also determines the phase of $\Delta=|\Delta|e^{i\theta}$, hence the dynamics of magnetization is directly related to the commutation relation $[\theta,S^{z}]=i$.

We can directly utilize the formalism in Ref.~\onlinecite{Nogueira04} to calculate the spin current by setting $m=(2U/3)\langle S^{z}\rangle=0$ therein. Consider that the magnetizations on the two sides have the same magnitude $|\Delta_{L}|=|\Delta_{R}|=|\Delta|$ but a difference in direction $\theta=\theta_{L}-\theta_{R}$. The spin supercurrent due to coherent tunneling of $\langle c_{\uparrow}^{\dag}c_{\downarrow}\rangle$ is 
\begin{eqnarray}
J_{s}&=&\frac{1}{2}\langle\dot{N}_{L\uparrow}-\dot{N}_{L\downarrow}\rangle=\frac{|T|^{2}}{2}S(0,|\Delta|)\sin\theta
\nonumber \\
&=&J_{s}^{0}\sin\theta=N\langle \dot{S}_{L}^{z}\rangle=-N\langle \dot{S}_{R}^{z}\rangle ,
\label{spin_current_FIF}
\end{eqnarray}
where $|T|^{2}$ represents the tunneling amplitude, and $N$ is total number of sites on either side of the junction. The function $S(a,b)$ satisfies $S(a,0)=0$ and $S(0,b)\neq 0$.


Applying a gate voltage on the thin insulating interface has two effects. Firstly, it changes the potential barrier at the interface, thus changing the tunneling amplitude $|T|^{2}$. Therefore one should replace $J_{s}^{0}\rightarrow J_{s}^{0}({\bf E})$. Secondly, the propagation of spinful $\langle c_{\uparrow}^{\dag}c_{\downarrow}\rangle$ picks up a phase that depends linearly on the electric field due to AC effect or Rashba SOC. From previous calculation for spin-FET, we anticipate that strong Rashba SOC is also necessary to experimentally observe the flux quantum in the spin Josephson effect. This leads us to consider a 2D version of FMM/I/FMM junction, with the insulating interface replaced by 2DEG (2DFMM/2DEG/2DFMM junction), as shown in Fig.\ref{fig:2DFMM2DEG2DFMM}(a). The Ginzburg-Landau (GL) free energy of the junction contains an AC phase 
\begin{eqnarray}
f=f_{n0}+\alpha|\psi|^{2}+\frac{\beta}{2}|\psi|^{4}-\frac{\hbar^{2}}{2m}|\left(\partial_{x}-ik_{0x}\right)\psi|^{2}\;,
\label{GL_FMMIFMM}
\end{eqnarray}
where $\alpha$ and $\beta$ are GL parameters, $m$ represents the effective mass, 
${\bf k}_{0}={\boldsymbol \mu}\times\left({\boldsymbol g}+\lambda{\bf E}\right)/\hbar$ and $k_{0x}$ is its component along the junction. Eq. (\ref{GL_FMMIFMM}) is consistent with the Hamiltonian of a spin current in an electric field \cite{Meier03}, with a fixed dipole moment ${\boldsymbol \mu}\parallel \langle{\bf S}_{R}\times{\bf S}_{L}\rangle$. The ${\boldsymbol g}$ and $\lambda$ again represent intrinsic and field dependence of Rashba SOC at the interface. The dimensionless quantity $g(x)=\psi(x)/\psi_{\infty}=\psi_{x}/(-\alpha/\beta)$ in the interface $0<x<L$, where the {\bf E} field is applied, satisfies the Laplace equation\cite{Tinkham96}
\begin{eqnarray}
\left(\partial_{x}-ik_{0x}\right)^{2}g=0\;.
\end{eqnarray}
The solution is 
\begin{eqnarray}
g(x)=(1-\frac{x}{L})e^{ik_{0x}x}+\frac{x}{L}e^{-ik_{0x}(L-x)+i\theta}\;,
\label{gx_interface}
\end{eqnarray}
such that it satisfies the boundary condition $g(0)=1$ and $g(L)=e^{i\theta}$. The GL free energy integrated over the interface is $\Delta F= l_{c}\hbar^{2}\psi_{\infty}^{2}/Lm\left[1-\cos\left(\theta-k_{0x}L\right)\right]$, where $l_{c}$ is the cross section length. The current from Eq.~(\ref{GL_FMMIFMM}) yields
\begin{eqnarray}
J_{s}=J_{s}^{0}({\bf E})\sin\left(\theta-\varphi_{0}-\varphi_{AC}\right)\;,
\label{Js_FET}
\end{eqnarray} 
where $\varphi_{0}={\boldsymbol \mu}\cdot\int {\boldsymbol g}\times d{\bf l}/\hbar$, and $\varphi_{AC}=\left(\lambda|{\boldsymbol\mu}|/\hbar\right){\boldsymbol{\hat\mu}}\cdot\int {\bf E}\times d{\bf l}=2\pi{\boldsymbol{\hat\mu}}\cdot{\boldsymbol\Phi}_{E}/\tilde{\Phi}_{E}^{0}$. The periodicity of Eq. (\ref{Js_FET}) again implies ${\boldsymbol\Phi}_{E}$ is quantized by $\tilde{\Phi}_{E}^{0}=h/\lambda|{\boldsymbol\mu}|$. Moreover, the current-phase relation of the Josephson spin current in this junction can be arbitrarily adjusted by the gate voltage, i.e., a $\varphi-$junction that shows high controllability by a gate voltage, which can have wide applications in, for instance, computation or data storage.

In the absence of Rashba SOC, $k_{0x}=0$, Eq.(\ref{gx_interface}) implies the magnetization profile in the 2DEG interface (green arrows in Fig.\ref{fig:2DFMM2DEG2DFMM}(c)) gradually rotates from $\langle{\bf S}_{L}\rangle$ to $\langle{\bf S}_{R}\rangle$, due to its proximity to the two FMMs. The gradient of this coplanar magnetization is the origin of the Josephson spin current that has a fixed polarization ${\boldsymbol \mu}\parallel \langle{\bf S}_{R}\times{\bf S}_{L}\rangle$. At large enough ${\bf E}$ field, the Rashba SOC gives additional rotation to the magnetization profile (red arrows in Fig.\ref{fig:2DFMM2DEG2DFMM}(d)). Hence the gate voltage changes the magnetization profile in the interface, which should be visible as a fringe pattern by polarization-sensitive probes such as optical Kerr effect\cite{Pechan05}, Lorentz transmission electron microscopy\cite{Graef00}, or magnetic transmission soft X-ray microscopy\cite{Fischer08}.

\section{Realization of He-McKellar-Wilkens effect by bilayer exciton condensate}

An intense investigation has been dedicated to the realization of HMW effect. Since there is no sizeable ${\bf d}$ for known point particles, existing proposals mainly focus on field induced electric dipole moments\cite{Wei95,Sato09}. Recently, interference effect due to electrically polarized $^{7}$Li ions has been observed in an atom interferometer\cite{Lepoutre12}, with an interferometer signal $I=I_{0}\left(1+{\cal V}\cos\left(\varphi_{p}+\varphi_{d}\right)\right)$ that depends on fringe visibility ${\cal V}$, perturbation phase $\varphi_{p}$, and diffraction phase $\varphi_{d}$. Following previous arguments from Sec. II to IV, it is tempting to directly associate the periodicity of $I$ to quantization of ${\boldsymbol\Phi}_{B}$. However, the complication of such electrically polarized atoms is that, by travelling through a region with both ${\bf E}$ and ${\bf B}$ field, the atom picks up a phase not only due to HMW effect, but also due to  AC, Zeeman, and Stark effect. Further treatment is necessary to extract the HMW phase. Therefore it is ambiguous to directly attribute the periodicity to flux quantization.

Here we propose that the bilayer exciton condensates observed in semiconductor heterostructures\cite{Spielman00,Kellogg04,Tutuc04,Eisenstein04,Tiemann08,Huang12,Sivan92,Seamons07,DasGupta08}  and in bilayer graphene\cite{Kim11,Kim12,Gorbachev12,Berman12} can also realize HMW effect. In the interferometer proposed below, no other effects contribute to interference, so that the interpretation of flux quantization due to HMW effect is straightforward. The generic system of this kind is a bilayer where electrons in one layer are strongly bound to holes on the other. At sufficiently low temperature, the bilayer excitons condense and their collective behavior can be described by a single superfluid-like wave function\cite{Su08}. Although the response of this bilayer condensate to external electromagnetic field has been discussed\cite{Moon95,Balatsky04}, below we emphasize that, in systems where the bilayer condensate travels in a closed trajectory, the interference due to combined AB effect of the electron and hole that make up the exciton can be regarded as HMW effect of the exciton\cite{Wei95}.

We first discuss the 2DEG/2DHG bilayer, where the condensate obviously has electric dipole moment, and address the 2DEG/2DEG bilayer later. Our starting point is a superfluid wave function that describes the exciton condensate composed of electrons and holes reside on different layers that are a distance ${\boldsymbol\delta}$ apart\cite{Su08}
\begin{eqnarray}
\Phi({\bf r})=e^{i\theta({\bf r})}|\Phi({\bf r})|=\langle c^{\dag}({\bf r}-{\boldsymbol\delta}/2)c({\bf r}+{\boldsymbol\delta}/2)\rangle\;,
\end{eqnarray}
where $c({\bf r})$ is the electron annihilation operator, ${\bf r}$ is the center of mass coordinate of the excitons, and the electrons and holes are assumed to have the same effective mass. If the magnitude of the condensate remains rigid, the energy of the system is\cite{Su08} 
\begin{eqnarray}
E=\int d^{2}{\bf r}\left[\frac{\hbar^2\rho_{s}}{2m^{\ast}}\left({\boldsymbol\nabla}\theta\right)^{2}-E_{J}\cos\theta\right]\;,
\label{E_bilayer_no_B}
\end{eqnarray}
where $m^{\ast}$ is the effective mass, $\rho_{s}$ is the density of the condensate, and $E_{J}\cos\theta$ represents the Josephson energy for interlayer tunneling. In the presence of a magnetic field, Eq. (\ref{E_bilayer_no_B}) is modified by\cite{Moon95,Balatsky04} 
\begin{eqnarray}
E&&=\int d^{2}{\bf r}\left\{\frac{\hbar^2\rho_{s}}{2m^{\ast}}\left[{\boldsymbol\nabla}\theta-\frac{e}{\hbar}{\bf A}\left({\bf r}+{\boldsymbol\delta}/2\right)\right.\right.
\nonumber \\
&&+\left.\frac{e}{\hbar}{\bf A}\left({\bf r}-{\boldsymbol\delta}/2\right)\right]^{2}-\left.E_{J}\cos\theta\right\}\;,
\end{eqnarray}
where $e=|e|$ is charge of the hole. After traveling around a closed loop, the AB effect of the electron along ${\bf r}_{e}={\bf r}-{\boldsymbol\delta}/2$ plus that of the hole along ${\bf r}_{h}={\bf r}+{\boldsymbol\delta}/2$ gives an extra phase to the condensate
\begin{eqnarray}
\varphi_{AB}^{h}+\varphi_{AB}^{e}&=&\frac{e}{\hbar}\oint_{C_{h}}{\bf A}\left({\bf r}_{h}\right)\cdot d{\bf r}_{h}-\frac{e}{\hbar}\oint_{C_{e}}{\bf A}\left({\bf r}_{e}\right)\cdot d{\bf r}_{e}
\nonumber \\
&=&\frac{e}{\hbar}\left(\Phi_{B}^{C_{h}}-\Phi_{B}^{C_{e}}\right)\;,
\label{AB_phase_bilayer}
\end{eqnarray}
where $C_{e}$ and $C_{h}$ represent the trajectory that electron and hole travel. For small $|{\boldsymbol \delta}|$, expanding Eq. (\ref{AB_phase_bilayer}) yields 
\begin{eqnarray}
\varphi_{AB}^{h}+\varphi_{AB}^{e}=\frac{e}{\hbar}
\oint_{C_{ex}}\left({\boldsymbol\delta}\cdot{\boldsymbol\nabla}\right){\bf A}({\bf r})\cdot d{\bf r}\;,
\end{eqnarray}
where $C_{ex}$ is the trajectory of the center of mass of the exciton. The following vector identity holds
\begin{eqnarray}
{\boldsymbol\nabla}\left({\boldsymbol\delta}\cdot{\bf A}\right)={\boldsymbol\delta}\times\left({\boldsymbol\nabla}\times{\bf A}\right)
+\left({\boldsymbol\delta}\cdot{\boldsymbol\nabla}\right){\bf A}\;,
\label{vector_identity}
\end{eqnarray}
since ${\bf A}\times\left({\boldsymbol\nabla}\times{\boldsymbol\delta}\right)=0$ and $\left({\bf A}\cdot{\boldsymbol\nabla}\right){\boldsymbol\delta}=0$ for a constant ${\boldsymbol\delta}$. Integrating the left hand side of Eq. (\ref{vector_identity}) over the closed loop $C_{ex}$ yields zero, so 
\begin{eqnarray}
\varphi_{AB}^{h}+\varphi_{AB}^{e}&=&
-\frac{e}{\hbar}\oint_{C_{ex}}\left[{\boldsymbol\delta}\times\left({\boldsymbol\nabla}\times{\bf A}\right)\right]\cdot d{\bf r}
\nonumber \\
&=&-\frac{1}{\hbar}\oint_{C_{ex}}\left({\bf d}\times{\bf B}\right)\cdot d{\bf r}=\varphi_{HMW}\;.
\label{bilayer_AB_HMW}
\end{eqnarray}
Therefore we prove that the sum of AB effects of the hole and the electron is equivalent to HMW effect of the exciton with fixed dipole moment. It should be noticed, though, that because the phase shift comes from the sum of two AB effects, a closed trajectory is necessary to observe it owing to its gauge field nature. This gauge versus non-gauge feature is an important difference between interference of a true point dipole and that of a physical dipole that consists of two charges.

Recently, Rademaker {\it et al.}\cite{Rademaker11} suggested to use bilayer exciton condensate in concentric rings(or concentric cylinders) to do quantum interference. In fact, this concentric rings geometry has been proposed sometime ago by Wei {\it et al.}\cite{Wei95} as a realization of HMW effect. This geometry does not belong to our analysis because ${\bf d}$ changes direction along the ring, so it is not fixed. Nevertheless, the phase gained by the dipole in the concentric rings is similar to that described by Eq.~(\ref{bilayer_AB_HMW}). We emphasize that because we consider fixed ${\bf d}$ in this work, the magnetic flux is the vector defined in Eq. (\ref{phase_AC_HMW}).

\begin{figure}[ht]
\begin{center}
\includegraphics[clip=true,width=0.9\columnwidth]{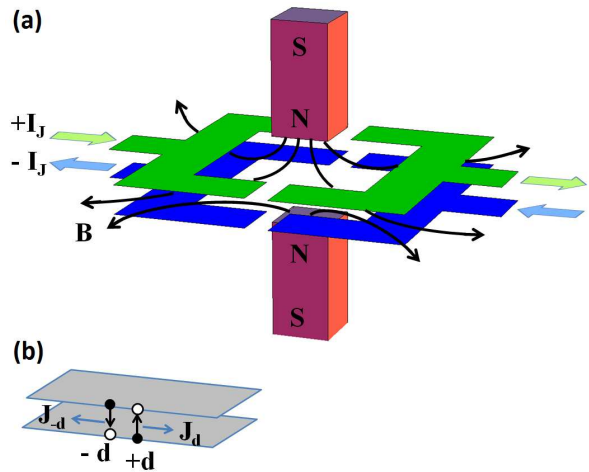}
\caption{ (color online) (a) Proposed dc SQUID-like device that uses HMW effect to create interference of bilayer exciton condensates. The green and blue sheets indicate the bilayer. The two oppositely placed magnets create a magnetic field that passes the space between the two layers, and cause a flux difference in the blue and green trajectories. The $\pm I_{J}$ indicates the counterflow experiment to measure the current. (b) Schematics of the two equivalent descriptions of the exciton condensate, one carries $+{\bf d}$ and the other $-{\bf d}$, in a quantum Hall bilayer at $\nu_{T}=1$. HMW effect causes the two condensates to have opposite Josephson currents, but the net dipole current is the same.  } 
\label{fig:bilayer_SQUID}
\end{center}
\end{figure}

From Eq. (\ref{AB_phase_bilayer}), it is clear that to observe interference of this condensate, the flux that passes the electron trajectory $\Phi_{B}^{C_{e}}$ and that passes the hole trajectory $\Phi_{B}^{C_{h}}$ must be different. A concrete design is by fabricating the Josephson junction of two sets of bilayers\cite{Park06,Wen96} into a dc SQUID-like geometry, as shown in Fig. \ref{fig:bilayer_SQUID}(a). There are many ways to create a difference in $\Phi_{B}^{C_{e}}$ and $\Phi_{B}^{C_{h}}$, such as placing the SQUID-like device is two oppositely placed magnets shown in Fig. \ref{fig:bilayer_SQUID}(a). As in a dc SQUID, the sum of Josephson dipole current on the two paths, labeled as $1$ and $2$, is
\begin{eqnarray}
J_{d}=\frac{J_{d}^{0}}{2}\left[\sin\left(\varphi_{0}-\varphi_{HMW}^{1}\right)+\sin\left(\varphi_{0}-\varphi_{HMW}^{2}\right)\right]\;,
\label{Josephson_dipole_current}
\end{eqnarray}
where $\varphi_{HMW}^{1}-\varphi_{HMW}^{2}=\oint\left({\bf B}\times{\bf d}\right)\cdot d{\bf r}/\hbar$. $\varphi_{0}$ is the intrinsic phase shift that can be tuned by an interlayer current\cite{Park06}. The Josephson dipole current should be measurable in the counterflow experiment\cite{Park06} as shown by the $\pm I_{J}$ in Fig. \ref{fig:bilayer_SQUID}(a). The typical interlayer distance is $|{\boldsymbol \delta}|\sim 10$nm, so $|{\bf d}|\sim 10^{-27}$Cm and $\tilde{\Phi}_{B}^{0}\sim 10^{-7}$mKg/Cs. Assuming the device in Fig. \ref{fig:bilayer_SQUID} of size mm can be managed to remain in quantum coherence, the magnetic field needed to observe one $\tilde{\Phi}_{B}^{0}$ is $\Delta B\sim 10^{-4}$T. $\Delta B$ is adjustable by tuning the interlayer distance, although one cannot achieve a precision higher than SQUIDs since the lattice constant sets up the limit for ${\bf d}$.

Exciton condensate has also been observed in 2DEG/2DEG quantum Hall bilayers with filling factor $\nu=1/2$ in each layer(total filling $\nu_{T}=1$). The physical picture for this case is that holes on layer 1 are bound to electrons on layer 2 and vice versa, when the interlayer distance ${\boldsymbol\delta}$ is small compared to magnetic length $l_{B}$ and the temperature is sufficiently low. The dipole moment of the whole system is zero, but as far as the dipole moment {\it of the condensate} is concerned, it can be considered as either a condensate that carries $+{\bf d}$, or a condensate that carries $-{\bf d}$ but lives on a different vacuum, as sketched in Fig. \ref{fig:bilayer_SQUID}(b). To see this, consider the bilayer state described by\cite{Fertig89}
\begin{eqnarray}
|\psi\rangle=\prod_{\bf k}\frac{1}{\sqrt{2}}\left(c_{{\bf k}\uparrow}^{\dag}+e^{i\varphi_{0}}c_{{\bf k}\downarrow}^{\dag}\right)|0\rangle\;.
\end{eqnarray}
Here $\uparrow$ and $\downarrow$ are layer indices (true spins are quenched by Zeeman effect), and ${\bf k}$ is the in-plane momentum. This state can be rearranged into a BCS-like form, in two different ways
\begin{eqnarray}
|\psi\rangle&=&\prod_{\bf k}\left(1+e^{i\varphi_{0}}c_{{\bf k}\downarrow}^{\dag}c_{{\bf k}\uparrow}\right)\prod_{{\bf k}^{\prime}}\frac{1}{\sqrt{2}}c_{{\bf k}^{\prime}\uparrow}^{\dag}|0\rangle
\nonumber \\
&=&\prod_{\bf k}\left(e^{-i\varphi_{0}}c_{{\bf k}\uparrow}^{\dag}c_{{\bf k}\downarrow}+1\right)\prod_{{\bf k}^{\prime}}\frac{e^{i\varphi_{0}}}{\sqrt{2}}c_{{\bf k}^{\prime}\downarrow}^{\dag}|0\rangle\;.
\end{eqnarray}
That is, either a condensate $e^{i\varphi_{0}}\langle c_{\downarrow}^{\dag}c_{\uparrow}\rangle$ that has $+{\bf d}$ and lives in the vacuum defined by $\prod_{{\bf k}^{\prime}}c_{{\bf k}^{\prime}\uparrow}^{\dag}|0\rangle/\sqrt{2}$, or $e^{-i\varphi_{0}}\langle c_{\uparrow}^{\dag}c_{\downarrow}\rangle$ that has $-{\bf d}$ but lives in the vacuum $\prod_{{\bf k}^{\prime}}e^{i\varphi_{0}}c_{{\bf k}^{\prime}\downarrow}^{\dag}|0\rangle/\sqrt{2}$. In fact, the linear combination of them is also possible. The two condensates have opposite initial phase $\pm\varphi_{0}$, and HMW effect causes the two condensates to pick up opposite phases $\pm\varphi_{HMW}$, so their Josephson currents, described by Eq. (\ref{Josephson_dipole_current}), are in opposite directions. But because they carry opposite dipole moments, the system has the same Josephson dipole current in either description. Therefore despite the whole system carries no dipole moment, because the condensate does carry dipole moment, one can also use $\nu_{T}=1$ 2DEG/2DEG bilayer to build the dc SQUID-like device in Fig. \ref{fig:bilayer_SQUID}(a). We remark that the magnetic field applied perpendicular to the plane to cause the Hall effect does not contribute to $\varphi_{HMW}$ because it is parallel to ${\bf d}$.

The device in Fig. \ref{fig:bilayer_SQUID}(a) can in turn be used to determine the experimental value of electric dipole moment ${\bf d}_{exp}$. Evidently, the theoretical value of electric dipole moment ${\bf d}_{theo}$ can be calculated from the interlayer distance. We expect that ${\bf d}_{exp}$ can be very different from ${\bf d}_{theo}$. Firstly, the difference can come from the wave function of particles trapped inside the potential well of 2DEG, which affects the mean distance between $+q$ and $-q$ that compose the dipole. Secondly, from Sec. III we know that because of confinement to lower dimensions, Rashba SOC dramatically enhances the relativistic coupling between magnetic dipoles and electric field, which makes quantization of ${\boldsymbol\Phi}_{E}$ experimentally accessible. We anticipate that the same phenomenon can happen for bilayer excitons, namely the confinement of the dipole in 2D may enhance its relativistic coupling to magnetic field. This means one may need to replace ${\bf B}\times{\bf d}\rightarrow\lambda_{d}{\bf B}\times{\bf d}$ in Eq. (\ref{momentum_AC_HMW}) with $\lambda_{d}>0$, which would reduce the flux quantum ${\tilde\Phi}_{B}^{0}$ and the field scale $\Delta B$ that the oscillation of dipole current can be seen. This anticipation, however, requires either experimental verification or further calculation, which is beyond the scope of this article. Nevertheless, we emphasize that if ${\bf d}_{exp}$ and ${\bf d}_{theo}$ significantly differ, it may imply certain new physics for the internal structure or relativistic effect of the electric dipole.



\section{Conclusions}

In summary, by comparing field-induced quantum interference effects of monopole and dipole moments, respectively, 
we clarify the principles that are universal to all interference effects and those that are unique to the interference of 
magnetic and electric
dipole moments. 
Central
is the general principle for flux quantization. By factoring the Berry phase into the part that depends only on the field and trajectory and the part that depends on the monopole or dipole moment, the flux is a scalar for particles carrying a monopole moment, but a vector for particles carrying a fixed dipole moment. This principle unifies all field-induced quantum interference devices that transport fixed monopole or dipole moments, including already known examples such as SQUID, spin-FET, and persistent charge or spin current in a mesoscopic ring. 

On the other hand, based on the non-gauge field character of the coupling between dipole moments and the external field, we demonstrate a unique feature of the interference effects of dipole moments: they can also take place in open trajectory devices, such as spin Josephson effect. 
In particular we show that because of the reduced flux quantum in systems with Rashba SOC, quantization of electric flux vector may become accessible to experiment. In addition, realization of the long sought HMW effect of electric dipole moments by bilayer exciton condensate is proposed, where quantization of magnetic flux vector should be easily observable. The device we propose may in turn be used to quantify the size of an electric dipole or its relativistic coupling to the magnetic field in the experimental condition. Finally, our calculation also indicates that AC and HMW effect, which were thought to cause only a small phase shift in atom interferometry, in fact also manifest themselves in a great number of solid state devices where the dipole currents are controlled by external fields.



We thank M. Sigrist, P. M. R. Brydon, O. P. Sushkov, H. Nakamura, W. Metzner, T. Kopp, and F. S. Nogueira for stimulating discussions.

\end{document}